\documentclass[ALICE,manyauthors]{cernphprep}
\usepackage[comma,square,numbers,sort&compress]{natbib}

\usepackage{lineno}
\usepackage{xspace}
\usepackage{hyperref}
\usepackage{xcolor}
\usepackage{amsmath}
\usepackage{upgreek}
\usepackage[T1]{fontenc}
\usepackage{orcidlink}

\begin{document}
%

\newcommand{\pp}           {pp\xspace}
\newcommand{\ppbar}        {\mbox{$\mathrm {p\overline{p}}$}\xspace}
\newcommand{\XeXe}         {\mbox{Xe--Xe}\xspace}
\newcommand{\PbPb}         {\mbox{Pb--Pb}\xspace}
\newcommand{\pA}           {\mbox{pA}\xspace}
\newcommand{\pPb}          {\mbox{p--Pb}\xspace}
\newcommand{\AuAu}         {\mbox{Au--Au}\xspace}
\newcommand{\dAu}          {\mbox{d--Au}\xspace}

\newcommand{\s}            {\ensuremath{\sqrt{s}}\xspace}
\newcommand{\snn}          {\ensuremath{\sqrt{s_{\mathrm{NN}}}}\xspace}
\newcommand{\pt}           {\ensuremath{p_{\rm T}}\xspace}
\newcommand{\meanpt}       {$\langle p_{\mathrm{T}}\rangle$\xspace}
\newcommand{\ycms}         {\ensuremath{y_{\rm CMS}}\xspace}
\newcommand{\ylab}         {\ensuremath{y_{\rm lab}}\xspace}
\newcommand{\etarange}[1]  {\mbox{$\left | \eta \right |~<~#1$}}
\newcommand{\yrange}[1]    {\mbox{$\left | y \right |~<~#1$}}
\newcommand{\dndy}         {\ensuremath{\mathrm{d}N_\mathrm{ch}/\mathrm{d}y}\xspace}
\newcommand{\dndeta}       {\ensuremath{\mathrm{d}N_\mathrm{ch}/\mathrm{d}\eta}\xspace}
\newcommand{\avdndeta}     {\ensuremath{\langle\dndeta\rangle}\xspace}
\newcommand{\dNdy}         {\ensuremath{\mathrm{d}N_\mathrm{ch}/\mathrm{d}y}\xspace}
\newcommand{\Npart}        {\ensuremath{N_\mathrm{part}}\xspace}
\newcommand{\Ncoll}        {\ensuremath{N_\mathrm{coll}}\xspace}
\newcommand{\dEdx}         {\ensuremath{\textrm{d}E/\textrm{d}x}\xspace}
\newcommand{\RpPb}         {\ensuremath{R_{\rm pPb}}\xspace}
\newcommand{\Raa}         {\ensuremath{R_{\rm AA}}\xspace}

\newcommand{\nineH}        {$\sqrt{s}~=~0.9$~Te\kern-.1emV\xspace}
\newcommand{\seven}        {$\sqrt{s}~=~7$~Te\kern-.1emV\xspace}
\newcommand{\twoH}         {$\sqrt{s}~=~0.2$~Te\kern-.1emV\xspace}
\newcommand{\twosevensix}  {$\sqrt{s}~=~2.76$~Te\kern-.1emV\xspace}
\newcommand{\five}         {$\sqrt{s}~=~5.02$~Te\kern-.1emV\xspace}
\newcommand{\twosevensixnn}{$\sqrt{s_{\mathrm{NN}}}~=~2.76$~Te\kern-.1emV\xspace}
\newcommand{\fivenn}       {$\sqrt{s_{\mathrm{NN}}}~=~5.02$~Te\kern-.1emV\xspace}
\newcommand{\LT}           {L{\'e}vy-Tsallis\xspace}
\newcommand{\GeVc}         {Ge\kern-.1emV/$c$\xspace}
\newcommand{\MeVc}         {Me\kern-.1emV/$c$\xspace}
\newcommand{\TeV}          {Te\kern-.1emV\xspace}
\newcommand{\GeV}          {Ge\kern-.1emV\xspace}
\newcommand{\MeV}          {Me\kern-.1emV\xspace}
\newcommand{\GeVmass}      {Ge\kern-.2emV/$c^2$\xspace}
\newcommand{\MeVmass}      {Me\kern-.2emV/$c^2$\xspace}
\newcommand{\lumi}         {\ensuremath{\mathcal{L}}\xspace}

\newcommand{\ITS}          {\rm{ITS}\xspace}
\newcommand{\TOF}          {\rm{TOF}\xspace}
\newcommand{\ZDC}          {\rm{ZDC}\xspace}
\newcommand{\ZDCs}         {\rm{ZDCs}\xspace}
\newcommand{\ZNA}          {\rm{ZNA}\xspace}
\newcommand{\ZNC}          {\rm{ZNC}\xspace}
\newcommand{\SPD}          {\rm{SPD}\xspace}
\newcommand{\SDD}          {\rm{SDD}\xspace}
\newcommand{\SSD}          {\rm{SSD}\xspace}
\newcommand{\TPC}          {\rm{TPC}\xspace}
\newcommand{\TRD}          {\rm{TRD}\xspace}
\newcommand{\VZERO}        {\rm{V0}\xspace}
\newcommand{\VZEROA}       {\rm{V0A}\xspace}
\newcommand{\VZEROC}       {\rm{V0C}\xspace}
\newcommand{\Vdecay} 	   {\ensuremath{V^{0}}\xspace}

\newcommand{\ee}           {\ensuremath{e^{+}e^{-}}} 
\newcommand{\pip}          {\ensuremath{\pi^{+}}\xspace}
\newcommand{\pim}          {\ensuremath{\pi^{-}}\xspace}
\newcommand{\kap}          {\ensuremath{\rm{K}^{+}}\xspace}
\newcommand{\kam}          {\ensuremath{\rm{K}^{-}}\xspace}
\newcommand{\pbar}         {\ensuremath{\rm\overline{p}}\xspace}
\newcommand{\kzero}        {\ensuremath{{\rm K}^{0}_{\rm{S}}}\xspace}
\newcommand{\lmb}          {\ensuremath{\Lambda}\xspace}
\newcommand{\almb}         {\ensuremath{\overline{\Lambda}}\xspace}
\newcommand{\Om}           {\ensuremath{\Omega^-}\xspace}
\newcommand{\Mo}           {\ensuremath{\overline{\Omega}^+}\xspace}
\newcommand{\X}            {\ensuremath{\Xi^-}\xspace}
\newcommand{\Ix}           {\ensuremath{\overline{\Xi}^+}\xspace}
\newcommand{\Xis}          {\ensuremath{\Xi^{\pm}}\xspace}
\newcommand{\Oms}          {\ensuremath{\Omega^{\pm}}\xspace}
\newcommand{\degree}       {\ensuremath{^{\rm o}}\xspace}
\newcommand{\Dzero}        {\ensuremath{\mathrm{D^{0}}}\xspace}
\newcommand{\DzeroJets}        {\ensuremath{\mathrm{D^{0}\mbox{-}jets}}\xspace}
\newcommand{\DzeroJet}        {\ensuremath{\mathrm{D^{0}\mbox{-}jet}}\xspace}

\begin{titlepage}
\PHyear{2024}       
\PHnumber{240}      
\PHdate{17 September}  

\title{Investigating charm quark energy loss in medium with the nuclear modification factor of D$^\textbf{0}$-tagged jets}

\ShortTitle{Nuclear modification factor of charm jets tagged with D$^0$ mesons}   

\Collaboration{ALICE Collaboration\thanks{See Appendix~\ref{app:collab} for the list of collaboration members}}
\ShortAuthor{ALICE Collaboration} 

\begin{abstract}
The nuclear modification factor $R_\mathrm{AA}$ of charm jets, identified by the presence of a D$^0$ meson among the jet constituents, has been measured for the first time in \PbPb collisions at a centre-of-mass energy per nucleon pair $\snn~=~5.02$ TeV with the ALICE detector at the LHC. The \Dzero\ mesons and their charge conjugates are reconstructed from the hadronic decay D$^0\to \mathrm K^{-}\pi^{+}$. Jets are reconstructed from D$^0$-meson candidates and charged particles using the anti-$k_\mathrm{T}$ algorithm with jet resolution parameter $R~=~0.3$, in the jet transverse momentum (\pt) range 5~$< p_\mathrm{T}^\mathrm{ch~jet}<$~50~GeV/$c$ and pseudorapidity $|\eta^\mathrm{ch~jet}| < 0.6$.  A hint of reduced suppression in the charm-jet \Raa is observed in comparison to inclusive jets in central \PbPb collisions with a significance of about 2$\sigma$ in 20~$< p_\mathrm{T}^\mathrm{ch~jet}<$~50 GeV/$c$, suggesting the in-medium energy loss to depend on both the difference between quark and gluon coupling strength (Casimir colour-charge effect) and quark mass (dead-cone effect).
The data are compared with model calculations that include mass effects in the in-medium energy loss. Several state-of-the-art models are consistent with the data, with the LIDO model providing the best description of the data in the common kinematic range of inclusive and \Dzero-tagged jets, highlighting the role of mass effects in interpreting the results.
\end{abstract}
\end{titlepage}

\setcounter{page}{2} 


\section{Introduction}

At the high energy density and temperature reached in ultrarelativistic heavy-ion collisions, the formation of a deconfined state of quarks and gluons, known as the quark--gluon plasma (QGP)~\cite{Heinz:2000bk,Busza:2018rrf,ALICE:2022wpn}, is predicted by quantum chromodynamic (QCD) calculations on the lattice~\cite{Busza:2018rrf,HotQCD:2018pds,Borsanyi:2020fev,Satz:2005hx}. The QGP is created and studied in high-energy heavy-ion collisions at the Relativistic Heavy Ion Collider (RHIC)~\cite{STAR:2005gfr,STAR:2006vcp, PHENIX:2005zfm, STAR:2002ggv, PHENIX:2004vcz} and the CERN Large Hadron Collider (LHC)~\cite{ALICE:2022wpn}.
Heavy quarks (charm and beauty), due to their large masses, are mainly produced in hard scattering processes that occur in the early stage of the collisions, on a shorter time scale compared to the QGP formation~\cite{Liu:2012ax,Andronic:2015wma}. 
Since thermal production and annihilation rates of heavy quarks in the strongly interacting medium are small, even at the high temperatures and densities reached in Pb--Pb collisions at the LHC, they are primarily produced in the initial stages of the collision. As a result, they experience the entire evolution of the system.

Jets, i.e.~collimated sprays of hadrons arising from the fragmentation of energetic partons produced in hard scattering processes, serve as effective probes for studying the microscopic interactions of partons with the QGP. Jets offer closer access to the kinematics of the original parton that initiates the shower, providing more direct information about its energy and direction with respect to single-hadron measurements. Furthermore, jets are easier to compare with QCD calculations and have reduced dependence on hadronization models, along with introducing a smaller bias from fragmentation~\cite{Djordjevic_2014}.
According to QCD~\cite{Tanabashi:2018oca,Buckley:2011ms, Dremin:2000ep}, the parton-shower evolution of a jet in vacuum depends on various factors, including the colour charge, characterized by the Casimir factor, and the quark mass, through the dead-cone effect. Notably, the dead-cone effect was recently directly measured for the first time by ALICE in pp collisions~\cite{ALICE:2021aqk}.

In heavy-ion collisions, high-energy partons traversing the QGP undergo additional energy loss through medium-induced gluon radiation and elastic scatterings~\cite{Aichelin:2012ww}. The extent of this energy loss depends on the parton colour charge and mass. For instance, gluons carrying a larger Casimir factor (3) compared to quarks (4/3) are expected to lose more energy. Similarly, heavy quarks are expected to experience reduced energy loss relative to light quarks, due in part to the suppression of small-angle radiation (dead-cone effect)~\cite{Armesto:2003jh,Zhang:2003wk,Djordjevic:2003zk}.
As a result, both the colour charge and mass of the initiating parton can contribute to the differences in energy loss observed between charm jets and inclusive jets, the latter being a mixture of quark- and gluon-initiated jets. 

The relative influence of these factors, particularly how much the quark mass and associated radiation suppression contribute to jet quenching, can be explored through measurements of the nuclear modification factor (\Raa) of jets in heavy-ion collisions.
This factor is defined as the ratio of the \pt-differential production yield in nucleus--nucleus collisions (d$N_\mathrm{AA}$/d\pt) and the production cross section in proton--proton collisions (d$\sigma_\mathrm{pp}$/d\pt) scaled by the average nuclear overlap function, $\langle T_\mathrm{AA} \rangle$
\begin{equation}
\label{Raa}
    \Raa = \frac{1}{ \left< T_{AA}\right>}\frac{\mathrm{d}^{2}N_\mathrm{Pb-Pb}/\mathrm{d}\eta^\mathrm{ch~jet} \mathrm{d}p^\mathrm{ch~jet}_\mathrm{T}}{\mathrm{d}^{2}\sigma_{\rm pp}/\mathrm{d}\eta^\mathrm{ch~jet} \mathrm{d}p^\mathrm{ch~jet}_\mathrm{T}},
\end{equation}
where $\langle T_\mathrm{AA} \rangle$ is estimated with Glauber model calculations~\cite{Miller:2007ri, dEnterria:2020dwq} and is related to the number of binary nucleon--nucleon collisions ($\langle N_\mathrm{coll} \rangle$) and the inelastic nucleon--nucleon cross section ($\sigma_{\mathrm{pp}}^\mathrm{inel}$) according to $\langle T_\mathrm{AA} \rangle = \langle N_\mathrm{coll} \rangle/\sigma_{\mathrm{pp}}^\mathrm{inel}$. The normalised ratio between the two collision systems can be used as an indicative that the number of jets in a given \pt interval is reduced in heavy-ion collisions in comparison to pp collisions, where the QGP formation is not expected. This suppression in heavy-ion collisions can be associated to parton energy loss.
The comparison of inclusive and heavy-quark jet \Raa can be used to observe whether the mass and colour charge effects are different for gluons, heavy and light quarks.

Several measurements at the LHC have investigated the energy loss of charm, beauty, and light quarks and gluons via the nuclear modification of fully reconstructed D mesons, non-prompt J/$\psi$, and charged hadrons in Pb--Pb collisions~\cite{ALICE:2018lyv,ALICE:2022tji,ALICE:2019nrq,CMS_2018474,PhysRevLett.109.222301,CMS_JHEP05(2012)063,2022137077,Aaboud2018, ALICE:2022iba, CMS:2017uoy}. The comparison with model predictions allowed for constraining the charm spatial diffusion coefficient and provided important insights into the role of radiative and collisional energy loss as well as into the hadronisation mechanism. Furthermore, the quark-mass dependence of parton energy loss was explored by comparing the \Raa of charm and beauty hadrons, which were either reconstructed directly or accessed through their decay products~\cite{ALICE:2018lyv,ALICE:2022tji, ALICE:2022xrg, CMS:2017uuv,ALICE:2022iba}. Models incorporating a quark-mass dependence of energy loss were able to describe the data.
On the jet side, the measurement of radial distributions of \Dzero\ mesons with respect to the jet axis, performed with the CMS detector~\cite{CMS:2019jis}, showed a hint of modification in \PbPb collisions for low-\pt \Dzero mesons. This suggests charm-quark diffusion in the medium, implying a potential influence of both mass and colour charge on partonic energy loss in heavy-ion collisions.
Recent measurements of the ATLAS Collaboration indicate a lower suppression of beauty jets~\cite{ATLAS:2022fgb} and of photon-tagged jets (mainly produced by light quarks)~\cite{ATLAS:2023iad} when compared with inclusive jets in central \PbPb collisions. This is evidence of a larger jet quenching in gluon jets, which are a dominant part of the inclusive jet sample at the LHC, in comparison to quark jets and it unequivocally highlights the role of the colour-charge effect.

The ALICE detector, with its excellent particle identification and tracking performance, provides the unique opportunity to tag charged-particle jets with reconstructed heavy-flavour hadrons at low \pt. Thus, it allows for the investigation of the low-jet-transverse-momentum region, where the mass effects are expected to be most pronounced.

In this paper, the yields of charm jets tagged with \Dzero mesons (\Dzero-tagged jets) and their nuclear modification factor in the most central (0--10\%) Pb--Pb events at centre-of-mass energy $\sqrt{s_{\mathrm{NN}}} = 5.02$ TeV are presented. The centrality is determined as in Ref.~\cite{ALICE-PUBLIC-2018-011}. This paper is organised as follows: Section~\ref{detectorsection} describes the ALICE detector and the utilised data sample; Sections~\ref{analysis} and~\ref{systematics} present details of the analysis strategy and the systematic uncertainty estimations;  Section~\ref{results} reports the results compared with inclusive jet measurements and model predictions. A summary is provided in Section~\ref{summary}.

\section{Detector and data sample}
\label{detectorsection}
The ALICE experimental apparatus~\cite{ALICE:2008ngc} consists of a central barrel (covering the pseudorapidity range $|\eta| <$ 0.9) embedded in a large solenoidal magnet that provides a magnetic field of 0.5 T, parallel to the beams axis. It also includes a forward muon spectrometer ($-4~< \eta <-$2.5) and a set of detectors at forward and backward rapidity used for triggering, background rejection, and event characterisation.

The central-barrel detectors employed in this paper for charged-particle reconstruction and identification at midrapidity are the Inner Tracking System (ITS)~\cite{ALICE:2010tia}, the Time Projection Chamber (TPC)~\cite{Alme:2010ke}, and the Time-Of-Flight detector (TOF)~\cite{Akindinov:2013tea}. The ITS consists of six layers of silicon detectors used for tracking charged particles and reconstructing primary and secondary vertices. The TPC serves as the main tracking detector and provides particle identification through the measurement of the particle specific energy loss (d$E$/d$x$) in the detector gas. Complementary particle identification information is obtained from the TOF, which measures the flight time of charged particles from the interaction point to the detector.

The results reported in this paper were obtained using the data sample collected during the 2018 LHC \PbPb run at centre-of-mass energy $\snn~=~5.02~$TeV. Events were selected using a minimum bias trigger provided by the V0 detectors~\cite{ALICE:2013axi}, which consist of two arrays of 32 scintillators each, covering the full azimuthal angle in the pseudorapidity ranges of $-3.7~< \eta < -1.7$ (V0C) and 2.8~$<~\eta~<$~5.1 (V0A). An additional trigger class, based on the online event selection provided by the V0 signal amplitude, was used during the data taking to enrich the sample of central collisions considered in this analysis. Events arising from the interactions of the beams with residual gas in the vacuum pipe were rejected offline using the timing information from the V0 detector and the Zero Degree Calorimeter ~\cite{PUDDU2009765}. Only the events with a primary vertex reconstructed within $\pm$10 cm from the nominal centre of the detector along the beam axis were considered in the analysis.
The centrality estimator is defined in terms of percentiles of the hadronic \PbPb cross section, using the sum of the V0 signal amplitudes, as described in detail in Ref.~\cite{ALICE:2010tia}. In the present analysis, only the 10\% most central collisions are used (0--10\% centrality class). The corresponding average nuclear overlap function is $\left \langle T_\mathrm{AA} \right \rangle = 23.26\pm0.17$~$\rm{mb}^{-1}$~\cite{ALICE:2015juo, ALICE-PUBLIC-2018-011} and the total number of analysed events is about $N_\mathrm{events}$~=~100 $\times$ 10$^6$, corresponding to an integrated luminosity of $L_\mathrm{int}$~=~130.5~$\pm$ 0.5~$\upmu$b$^{-1}$. The data used for the measurement of the pp reference in the calculation of the \Raa is the same as used in Ref.~\cite{ALICE:2022mur}.

The Monte Carlo samples used for the corrections in this analysis were generated by simulating pp collisions containing a $\mathrm{c\overline{c}}$ or a $\mathrm{b\overline{b}}$ pair with PYTHIA~8~\cite{Sjostrand:2007gs} (Monash 2013 tune~\cite{Skands:2014pea}). The charged-particle multiplicity and detector occupancy observed in data~\cite{ALICE:2012xs} were simulated superimposing an underlying event generated with HIJING 1.36~\cite{PhysRevD.44.3501}. The generated-level particles from these simulations were passed through a particle transport and detector-response simulation of the entire ALICE apparatus based on GEANT3~\cite{Brun:1082634}.

\section{Analysis Strategy}
\label{analysis}
The analysis closely follows the procedure described in detail in the ALICE studies of charm jets tagged with \Dzero mesons~\cite{ALICE:2022mur}. It consists of three main parts: (i) reconstruction of jets tagged by the presence of a D$^{0}$ meson; (ii) extraction of the raw yields of \Dzero-tagged jets; (iii) correction for the reconstruction efficiency of the \Dzero-tagged jets, subtraction of the feed-down contribution from \Dzero mesons coming from beauty decays, and correction for the detector-related effects and underlying-event fluctuations.

A major challenge in jet reconstruction in central heavy-ion collisions compared to pp collisions is the fluctuation of the underlying event, which is a background to the reconstructed jet. These fluctuations significantly affect the jet transverse momentum. To account for these background fluctuations and detector effects, the data were corrected using a deconvolution procedure (unfolding). This correction involved constructing a response matrix that maps the D$^{0}$-tagged-jet \pt at the generated level, simulated using PYTHIA 8 for pp collisions, to the detector-level spectrum. The matching between generated and detector-level D$^{0}$-tagged jets was achieved by requiring the identification of the same D$^{0}$ meson among their constituents. To ensure a realistic representation of the heavy-ion environment, detector-level particles were embedded into real Pb--Pb data to simulate the background accurately.

\subsection{\texorpdfstring{D$^\textbf{0}$}{D0}-meson selection and jet reconstruction}
The \Dzero and $\mathrm{\overline{\Dzero}}$ mesons were reconstructed via the hadronic decay channel D$^{0}~\to~\kam\pip$ (and its charge conjugate) with branching ratio BR = 3.950 $\pm$ 0.031\%~\cite{Tanabashi:2018oca}. D$^{0}$ mesons produced either directly from charm-quark fragmentation or from decays of directly-produced excited charm hadron states are referred to as prompt D$^{0}$ mesons, while those originating from decays of beauty hadrons are termed non-prompt D$^{0}$ mesons. The selection strategy, described in detail in Refs.~\cite{ALICE:2021rxa} and~\cite{ALICE:2018lyv}, exploited the displaced topology of the decay and made use of the particle identification capabilities of the TPC and TOF to identify the decay particles of the D$^{0}$ mesons in the pseudorapidity interval $|\eta_{\mathrm{D}}| < 0.9$.

Charm-jet reconstruction was performed using a track-based procedure with the FastJet~\cite{Cacciari:2011ma} anti-$k_\mathrm{T}$ clustering algorithm~\cite{Cacciari:2008gp} with resolution parameter $R=0.3$. This parameter was chosen in order to keep under control the jet background fluctuations, which increase for large $R$, while the jet's characteristics are not completely determined by the jet core that spans over smaller $R$. The jet reconstruction was carried out using the \pt-recombination scheme~\cite{Salam:2010nqg}. Charged particles were required to have \pt $>$ 150 \MeVc and $|\eta| <$ 0.9.
To ensure that the entire jet was contained within the detector acceptance, jets were required to have their axes within the pseudorapidity range of $|\eta| < 0.9 - R$.

At low momenta, the \Dzero decay products may be emitted at angles larger than the defined jet cone size. To address this issue and ensure that the pion and kaon from the D$^{0}$ decay were assigned to the same jet, they were removed from the set of charged-particle tracks before the jet reconstruction and their four-momenta were replaced by that of the D$^{0}$ candidate. A charm jet was identified when a D$^{0}$-meson candidate in the transverse momentum range 3$~<p_{\mathrm{T,D}}<~$36 \GeVc was found among its constituents. This procedure was repeated for each D$^{0}$-meson candidate in the event, where the decay products of only one candidate were replaced at a time.

Measuring jets in heavy-ion collisions down to low transverse momentum, such as 5 \GeVc, is particularly challenging due to the large background from the underlying event and its fluctuations~\cite{ALICE:2012nbx}. At low \pt, a large fraction of the reconstructed jets is not related to a hard scattering process. This background source dominates jets at low-\pt (below 20 \GeVc). Moreover, independently of the \pt interval, all reconstructed jets can include tracks from the underlying event, i.e.~particles not originating from the fragmentation of the hard-scattered parton that initiated the jet. To account for the background coming from the underlying event, the area-based method was employed, following the approach described in Refs.~\cite{ALICE:2013dpt, ALICE:2015mjv}. This method estimates the average additive contribution to the jet momentum on a jet-by-jet basis. The underlying background momentum density, $\rho^\mathrm{ch}$, was estimated event-by-event using the median of $p^{\mathrm{ch~jet}}_{\mathrm{T,~raw}}/A_{\mathrm{raw}}^{\mathrm{ch~jet}}$, where $p^{\mathrm{ch~jet}}_{\mathrm{T,~raw}}$ is the uncorrected jet transverse momentum and $A_{\mathrm{raw}}^\mathrm{ch~jet}$ is the area of jets reconstructed with the $k_\mathrm{T}$ algorithm~\cite{Ellis:1993tq}. The two leading jets were removed from the calculation of $\rho^\mathrm{ch}$. The signal anti-$k_\mathrm{T}$ jets were then corrected by subtracting the median of the jet transverse momentum density multiplied by the jet area: $p^\mathrm{ch~jet}_\mathrm{T,~corr}=p^{\mathrm{ch~jet}}_{\mathrm{T,~raw}} - \rho^\mathrm{ch}A_\mathrm{raw}^\mathrm{ch~jet}$.  The reported kinematic range is 5$~<p_{\mathrm{T,~raw}}^{\mathrm{ch~jet}}<~$50 \GeVc.

\subsection{Extraction of \texorpdfstring{D$^\textbf{0}$}{D0}-tagged jet raw yields}
\label{YieldExtraction}
The raw yield of \Dzero-tagged jets was determined through an invariant-mass analysis of the D$^{0}$-meson candidates used for tagging the charm-jet candidates. The analysis was performed in different intervals of D$^{0}$-meson transverse momentum ranging from 3 to 36 \GeVc. For each $p_\mathrm{T,D}$ interval, the invariant-mass distribution of the D$^{0}$-meson candidates was fitted with a function consisting of a Gaussian function to represent the signal and an exponential function to account for the combinatorial background~\cite{ALICE:2021rxa}.

When the two tracks forming a \Dzero candidate are compatible with both the kaon and pion hypothesis, the candidate can be identified both as a \Dzero and as a $\overline{\Dzero}$, leading to an irreducible correlated background, which is referred to as reflections~\cite{ALICE:2021rxa, ALICE:2019nxm}.
The contribution of residual \Dzero-meson reflections, which were not rejected by particle identification, was considered in the fit by including a template composed of the sum of two Gaussian functions. The centroids and widths of these Gaussian functions were fixed to the values based on a fit to the invariant mass distributions of reflections derived from the simulation. The ratio between the reflected signal and the D$^{0}$-meson yield was also fixed to the value obtained from the simulations.

The yield of D$^{0}$-tagged-jet candidates was determined by dividing the corresponding D$^{0}$ invariant-mass range in two sub-samples within each $p_\mathrm{T,D}$ interval: a $\textit{peak region}$ corresponding to candidates with $|m_\mathrm{inv} - m_\mathrm{fit}|< 2\sigma_\mathrm{fit}$ (where $m_\mathrm{fit}$ and $\sigma_\mathrm{fit}$ are the mean and width of the Gaussian component of the fit, respectively) and a $\textit{sideband region}$ consisting of candidates with 4$\sigma_\mathrm{fit} <~|m_\mathrm{inv} -m_\mathrm{fit}|< 8\sigma_\mathrm{fit}$. The jet \pt distribution associated with the D-meson candidates in the sideband region was normalised to the integral of the combinatorial background in the peak region, consisting of the exponential and reflections components, and then subtracted from the jet-\pt distribution in the peak region to obtain the raw yield of \Dzero-tagged jets.

\subsection{Corrections of \texorpdfstring{D$^\textbf{0}$}{D0}-tagged-jet raw yields}
\label{YieldsCorrection}
The raw yields of D$^{0}$-tagged jets extracted using the invariant mass method required corrections to account for the limited detector acceptance and the reconstruction efficiency of D$^{0}$ mesons. These corrections are described in detail in Refs.~\cite{ALICE:2022mur, ALICE:2019cbr}. After the removal of non-prompt \Dzero meson contributions, the raw yield is unfolded to correct for the detector finite resolution and jet background fluctuations.

It is worth noting that several ingredients of the correction procedure are very similar to those used in the pp reference analysis~\cite{ALICE:2022mur} and in the D$^{0}$-meson measurements in Pb--Pb collisions~\cite{ALICE:2021rxa} (e.g.\ acceptance$\times$efficiency corrections for prompt and non-prompt contributions, and invariant-mass fits). For this reason, the reader is referred to the corresponding references for a detailed description of these procedures and related figures.

First, the raw yields obtained in different intervals of $p_\mathrm{T,D}$ were corrected for the product of the acceptance and the reconstruction efficiency of prompt D$^{0}$ mesons associated to jets that pass the acceptance and $p_\mathrm{T}^\mathrm{ch~jet}$ selection, $\epsilon^{\rm{c\rightarrow\Dzero}}(p_{\rm{T,D}})$. This correction factor took into account the probability of detecting and reconstructing jets containing prompt D$^{0}$ mesons within the specified $p_\mathrm{T,D}$ intervals and the jet cone radius.

A fraction of the reconstructed D$^{0}$-meson-tagged jets originates from the fragmentation of beauty quarks, where a beauty hadron decays into a D$^{0}$ meson. The contribution from beauty jets was estimated using generated templates of non-prompt \Dzero-jet \pt distributions obtained with NLO pQCD calculations of POWHEG~\cite{POWHEG_1,POWHEG_2, POWHEG_3, POWHEG_4} coupled to the PYTHIA~6~\cite{Sj_strand_2006} parton shower. The templates were generated with specific configurations in POWHEG, such as a b-quark mass $m_\mathrm{b}=4.75$ GeV/\textit{c}$^2$, the renormalisation ($\mu_\mathrm{R}$) and factorisation ($\mu_\mathrm{F}$) scales set to the quark transverse mass $\mu_\mathrm{R} = \mu_\mathrm{F} = \sqrt{m_\mathrm{b}^2 + p_\mathrm{T}^2}$, and the use of parton distribution functions (PDF) from the CT10NLO~\cite{Lai:2010vv} set using the LHAPDF6~\cite{Buckley:2014ana} interpolator.

The non-prompt-D$^{0}$-jet yield estimated by the simulation ($N^{\rm b\rightarrow\Dzero}_{\rm{POWHEG}}$) was weighted by the ratio of the efficiencies of non-prompt and prompt D$^{0}$-tagged jets ($\epsilon^{\rm{b\rightarrow\Dzero}}(p_{\rm{T,D}})/\epsilon^{\rm{c\rightarrow\Dzero}}(p_{\rm{T,D}}))$ in order to be comparable with $\epsilon^{\rm{c\rightarrow\Dzero}}$-scaled inclusive D$^{0}$-tagged-jet distributions from data. As a next step, the simulated non-prompt D$^{0}$-tagged-jet distribution was smeared with the non-prompt response matrix (RM$_\mathrm{b\to D^{0}}$) in order to take into account the effect of the detector on the jet momentum resolution. The spectrum was also corrected for the kinematic efficiency, which accounts for the limited transverse momentum ranges of the response matrix in the generated ($\epsilon_\mathrm{kine}^\mathrm{gen}$) and reconstructed ($\epsilon_\mathrm{kine}^\mathrm{rec}$) axes. Additionally, the measured nuclear modification factor of non-prompt-D$^{0}$ mesons ($R_\mathrm{AA}^\mathrm{b\to D^{0}}$)~\cite{ALICE:2022tji} was used to weight the non-prompt \Dzero-jet \pt spectrum.
This assumes that the non-prompt \Dzero-jet \Raa is dominated by the non-prompt \Dzero. This assumption is supported by the expected hard fragmentation of the bottom quark.
As a result, the yield of prompt-D$^{0}$-tagged jets, $N^\mathrm{\rm c}$, was determined as follows

\begin{equation}
\label{Nc_corr}
N^\mathrm{c}(p_\mathrm{T}^\mathrm{ch~jet}) = N^\mathrm{c+b}(p_\mathrm{T}^\mathrm{ch~jet}) - N^{\rm b}(p_\mathrm{T}^\mathrm{ch~jet}),
\end{equation}

where the contribution of non-prompt D$^{0}$-tagged jet, $N^\mathrm{\rm b}$, was estimated using the following formula

\begin{equation}
\label{Nb}
N^{\mathrm{b}}(p_\mathrm{T}^\mathrm{ch~jet}) = \frac{1}{\epsilon^{\rm{rec}}_{\rm{kine}}}
\left[\sum_{p_{\rm T,D}}\mathrm{RM}_{\mathrm{b\rightarrow\Dzero}}(p_\mathrm{T}^\mathrm{ch~jet}, p_{\rm{T,D}})\otimes \sum_{p_{\rm T,D}} \frac{\epsilon^{\mathrm{b\rightarrow\Dzero}}(p_{\mathrm{T,D}})}{\epsilon^{\rm{c\rightarrow\Dzero}}(p_{\rm{T,D}})} R_{AA}^{\rm{b\rightarrow\Dzero}}(p_{\rm T,D}) \epsilon^{\rm{gen}}_{\rm{kine}} N^{\rm b\rightarrow\Dzero}_{\rm{POWHEG}}(p_\mathrm{T}^\mathrm{ch~jet}, p_{\rm{T,D}})
\right].
\end{equation}
\\
The contribution of non-prompt \Dzero-tagged jets varies between 20\% and 24\% of the inclusive yields depending on the jet \pt.

Finally, the measured D$^{0}$-tagged-jet yield was unfolded with an iterative method based on Bayes' theorem~\cite{DAgostini:2010hil} as implemented in the RooUnfold package~\cite{RooUnfold}. The chosen number of iterations was eight, based on optimal convergence.
Both the prompt and non-prompt-D$^{0}$-jet response matrices were corrected for the kinematic efficiency.

To verify the stability of the unfolding and the choice of the number of iterations, the unfolded spectra were folded back and compared to the original data. The ratio showed to be consistent with unity within uncertainties.

\section{Systematic uncertainties}
\label{systematics}

The systematic uncertainty associated to the measurement has several independent sources. These sources were identified as coming from: (i) yield extraction, (ii) D$^{0}$-tagged-jet reconstruction efficiency, (iii) feed-down subtraction, (iv) tracking efficiency, (v) unfolding, and (vi) branching ratio.

The uncertainty on the yield extraction procedure, described in Section~\ref{YieldExtraction}, was evaluated by varying the fit approach. In particular, the fit was repeated using different functions to describe the background (the default exponential was replaced by first and second order polynomial functions), varying the fit range, fixing the mean of the Gaussian term describing the signal peak to the nominal value of the D$^{0}$-meson mass or fixing the Gaussian width to the value obtained from Monte Carlo studies. Moreover, the signal range was varied between 2 to 3 standard deviations of the peak width, while the sideband extraction range $|m_\mathrm{inv} - m_\mathrm{fit}|$ was varied across different ranges of standard deviations. The systematic uncertainty was estimated as the standard deviation of all variations relative to the default parameter set. The uncertainties obtained with this procedure increase from 5\% at low jet transverse momentum to 6\% at higher jet \pt.

The possible differences in the topological variable distributions between Monte Carlo and data can affect the D$^{0}$-tagged-jet reconstruction efficiency. The related systematic uncertainty was evaluated by extracting the raw-jet \pt\ spectrum with tighter and looser topological selections of the D$^{0}$-tagged-jet candidates, with a corresponding variation of the D-meson reconstruction efficiencies larger than $\pm$10--25\%. The corresponding systematic uncertainty varies from 5\% at the low jet transverse momentum up to 14\% in the highest \pt interval (30--50 \GeVc).

The uncertainty on the subtraction of the beauty feed-down contribution was quantified by varying the factorisation and renormalisation scales, the beauty-quark mass, and the PDF in the generated POWHEG + PYTHIA~6 templates of non-prompt-D$^{0}$ jets~\cite{Cacciari_2012}. The largest deviations among all variations were used as systematic uncertainties, resulting in a value of 7--8\% depending on the jet \pt. An additional contribution to this uncertainty comes from the systematic uncertainty related to the assumption on the nuclear modification factor $R_\mathrm{AA}^\mathrm{b\to D^{0}}$. It was estimated by using the upper and lower bands of the uncertainties of the non-prompt-\Dzero nuclear modification factor~\cite{ALICE:2022tji} and recalculating the non-prompt \Dzero-jet yield $N^\mathrm{b}$.
The difference between the variations and the default spectrum was used as systematic uncertainty. The uncertainty ranges from 6\% to 8\% in jet \pt kinematic interval.

The uncertainties on the track reconstruction efficiency can impact the jet momentum resolution and \Dzero-tagged-jet reconstruction efficiency. The systematic uncertainty on the reconstruction efficiency of a single track was estimated to be 4\%~\cite{PhysRevC.101.034911}. To evaluate its impact, a response matrix was built by randomly rejecting 4\% of the reconstructed tracks (artificially decreasing the tracking efficiency). This response matrix was used to unfold the measurement and the difference with respect to the results obtained with the default response matrix was considered as systematic uncertainty. This uncertainty varies from 0.5\% in the 5--6 \GeVc jet \pt range up to 10\% in the 30--50 \GeVc jet-\pt interval.

The systematic uncertainties due to the unfolding procedure were evaluated by studying three independent aspects of the Bayesian unfolding method. First, the number of iterations as the regularisation parameter was varied by $\pm$1 iteration. These two variations were compared to the unfolded distribution obtained using the default parameter and the largest deviations were used as systematic uncertainty. Second, the shape of the prior spectrum of the \Dzero-tagged-jet transverse-momentum distribution in the response matrix was also varied. The default shape was taken from POWHEG+PYTHIA 6 simulations and the variation used was from PYTHIA 8. The observed difference was adopted as systematic uncertainty.
Third, the lower limit of the kinematic range of the measurement was also varied and, after unfolding, the spectrum was compared to the default range and the relative deviation was used as systematic uncertainty. The combined systematic uncertainties from unfolding range from 8\% at low jet \pt to 16\% in the 30--50 \GeVc interval.

The systematic uncertainties arising from the sources that affect the raw jet-\pt spectrum (such as yield extraction, kinematic and topological selections, and feed-down subtraction) were propagated to the final \Dzero-tagged-jet yield by unfolding the upper and lower bands of the combined systematic uncertainty. These were then combined in quadrature with the systematic uncertainties affecting the correction of the reconstructed jet momentum (including single-particle tracking efficiency and unfolding). This resulted in a combined systematic uncertainties for \Dzero-tagged jets that varies from 15\% in the 5--6 \GeVc jet \pt range up to 52\% in the 30--50 \GeVc interval.

In the evaluation of the nuclear modification factor, the systematic uncertainties associated with the \Dzero-tagged-jet yield in Pb--Pb collisions and the reference cross section in pp collisions were treated as uncorrelated, with the exception of the uncertainty related to the branching ratio (BR), which cancels out in the ratio, and the feed-down that was considered as partially correlated between pp and Pb--Pb collisions.
The contributions of the uncertainties on the luminosity determination in pp collisions, and the $\langle T_\mathrm{AA}\rangle$ estimated with the Glauber model are common across all the transverse-momentum intervals and therefore contribute to a normalisation uncertainty on the \Raa, which is shown separately from the other sources when displaying the results. A summary of the \Dzero-jet \Raa relative systematic uncertainties is displayed in table~\ref{t_SysUnc} with the contributions from Pb--Pb and pp collisions.

\begin{table}
\caption{Relative systematic uncertainties for selected $p_\mathrm{T}^\mathrm{ch~jet}$ intervals of the \Dzero-jet spectra in Pb--Pb and pp collisions. In Pb--Pb collisions, the first set of uncertainties is associated with the \Dzero-tagged-jet $p_\mathrm{T}^\mathrm{ch~jet}$ spectrum before unfolding. The second set (with different binning) is associated with the $p_\mathrm{T}^\mathrm{ch~jet}$ spectrum after unfolding.}
\label{t_SysUnc}
\begin{center}
\begin{tabular}{|lccc|}
\hline
\multicolumn{4}{|c|}{Pb--Pb collisions}\\
\hline
Source & \multicolumn{3}{c|}{Uncertainty (\%)}\\
\hline
$p_\mathrm{T}^\mathrm{ch~jet}$ (\GeVc) & [5,8] & [15,20] & [30,50]\\
\hline
 Raw yield extraction & 6 & 5 & 5\\
 \cline{1-4}
 B feed-down & 7 & 6 & 8\\
 \cline{1-4}
 Topological selection & 6 & 6 & 14\\
 \hline
$p_\mathrm{T}^\mathrm{ch~jet}$ (\GeVc) & [5,6] & [14,20] & [30,50]\\
 \hline
 Propagated & 13 & 16 & 48\\
 \cline{1-4}
 Unfolding & 8 & 15 & 16\\
 \cline{1-4}
 Tracking eff. & 0.5 & 3 & 10\\
 \cline{1-4}
 Total Pb--Pb & 15 & 23 & 52\\
 \hline \hline
 \multicolumn{4}{|c|}{pp collisions}\\
 \hline
 Source & \multicolumn{3}{c|}{Uncertainty (\%)}\\
\hline
$p_\mathrm{T}^\mathrm{ch~jet}$ (\GeVc) & [5,6] & [14,20] & [30,50]\\
\hline \hline
 Raw yield extraction & 3 & 3 & 7\\
 \cline{1-4}
 B feed-down & 6 & 12 & 14\\
 \cline{1-4}
 Topological selection & 1 & 6 & 6\\
 \hline
 Unfolding & 6 & 4 & 9\\
 \cline{1-4}
 Tracking eff. & 3 & 4 & 8\\
 \cline{1-4}
 Total pp & 10 & 15 & 21\\
 \hline \hline
 Total \Raa & 19 & 28 & 56\\
 \hline
\end{tabular}
\end{center}
\end{table}

\section{Results and discussion}
\label{results}

The \pt differential yields of charm jets tagged with prompt \Dzero mesons in central Pb--Pb collisions at \snn~=~5.02 TeV were calculated using the following equation~(\ref{yields}), with the measured yields corrected according to the procedure described in Section~\ref{analysis}:

\begin{equation}
   \frac{\mathrm{d^2}N}{\mathrm{d}\eta^\mathrm{ch~jet}\mathrm{d}p_\mathrm{T}^\mathrm{ch~jet}}= \frac{1}{N_{\mathrm{events}}}\frac{1}{\mathrm{BR}}\frac{N(p_\mathrm{T}^\mathrm{ch~jet})}{\Delta\eta^\mathrm{ch~jet} \Delta p_\mathrm{T}^\mathrm{ch~jet}},
   \label{yields}
\end{equation}

where $N_\mathrm{events}$ is the number of events, BR is the branching ratio of the considered \Dzero\ decay channel, $N(p_\mathrm{T}^\mathrm{ch~jet})$ is the measured yield in each interval of transverse momentum, $\Delta\eta^\mathrm{ch~jet}$ is the pseudorapidity interval of \Dzero-tagged jets and $\Delta p_\mathrm{T}^\mathrm{ch~jet}$ is the width of the \pt interval.

\begin{figure}[hbp]
\centering
\includegraphics[width=0.6\textwidth]{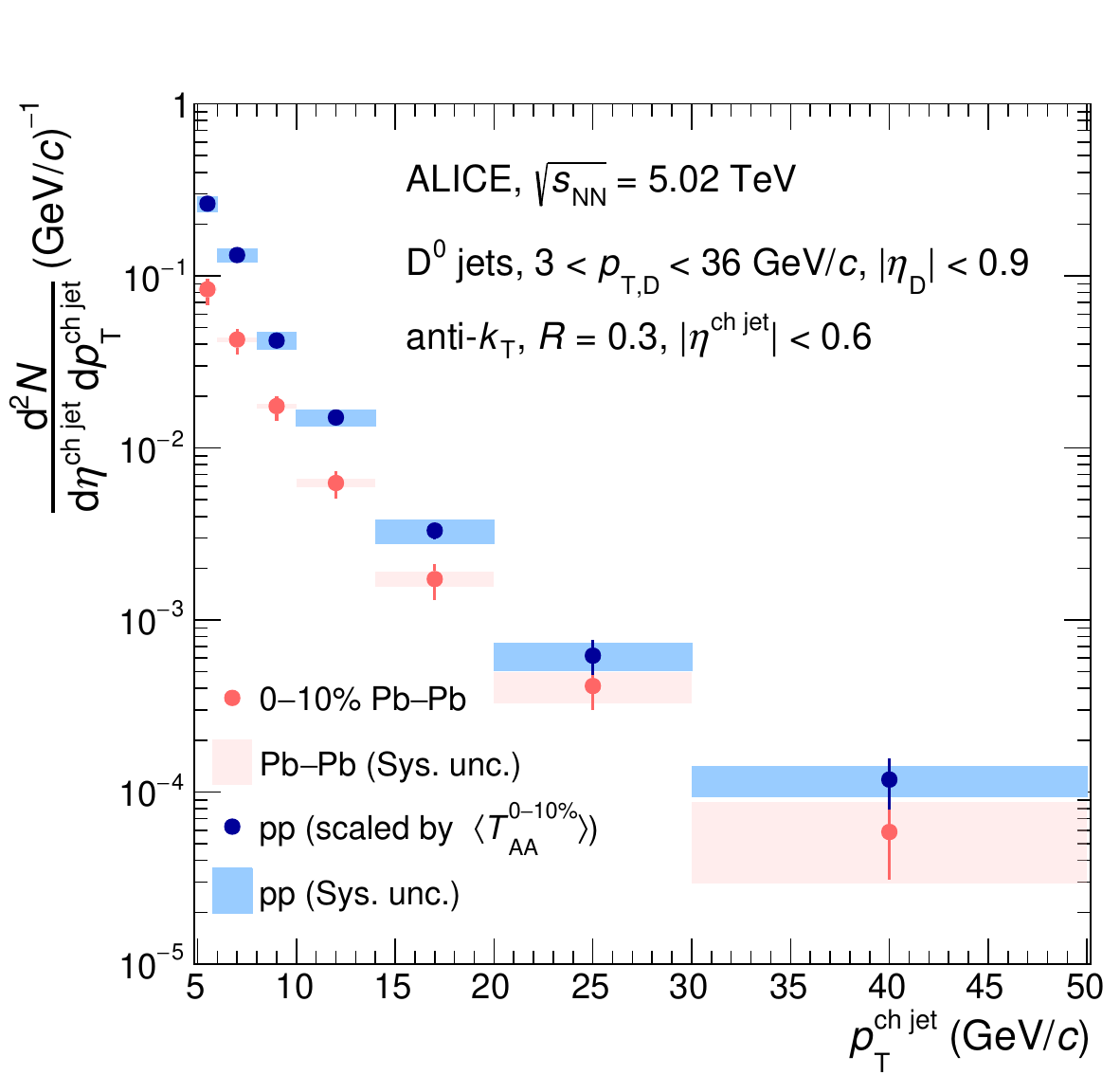}
\caption{Transverse-momentum differential yields of charm jets tagged with prompt \Dzero mesons in central (0--10\%) Pb--Pb collisions at $\snn\ = 5.02$~TeV (red circles) and cross section in pp collisions at $\s\ = 5.02$~TeV (blue squares) scaled by the nuclear overlap function $\left \langle T_\mathrm{AA} \right \rangle$ for the considered centrality interval.The vertical lines and boxes represent the statistical and systematic uncertainties, respectively.}
\label{fig:CharmJetYields}
\end{figure}

Figure~\ref{fig:CharmJetYields} displays the measured yields of charm jets tagged with a prompt \Dzero\ meson in the transverse momentum interval $5~<~p_\mathrm{T}^\mathrm{ch~jet}~<~50$~\GeVc for central (0--10\%) Pb--Pb collisions at a centre-of-mass energy of $\snn=5.02$~TeV, along with the reference yields from pp collisions~\cite{ALICE:2022mur}. The reference yields in pp collisions were computed as the product of the average nuclear overlap function $\langle T_\mathrm{AA}\rangle$ and the \Dzero-tagged-jet \pt-differential cross section $\mathrm d\sigma_\mathrm{pp}/\mathrm d p_\mathrm{T}$. In both collision systems, the charm jets were required to contain a prompt \Dzero\ as one of their constituents, reconstructed in the kinematic interval $3 < p_\mathrm{T,D} < 36$ \GeVc. This requirement allowed for the reconstruction of \Dzero-tagged jets with low transverse momentum. The measured yields in Pb--Pb collisions exhibit a clear suppression of the \Dzero-tagged-jet production yield compared to the reference yields in \pp\ collisions.
The nuclear modification factor (\Raa) of prompt-\Dzero-tagged jets was computed, based on Eq.~(\ref{Raa}), using the transverse momentum differential yield measured in Pb--Pb collisions and the pp reference at the same centre-of-mass energy~\cite{ALICE:2022mur}.

\begin{figure}[tbp]
\centering
\includegraphics[width=0.6\textwidth]{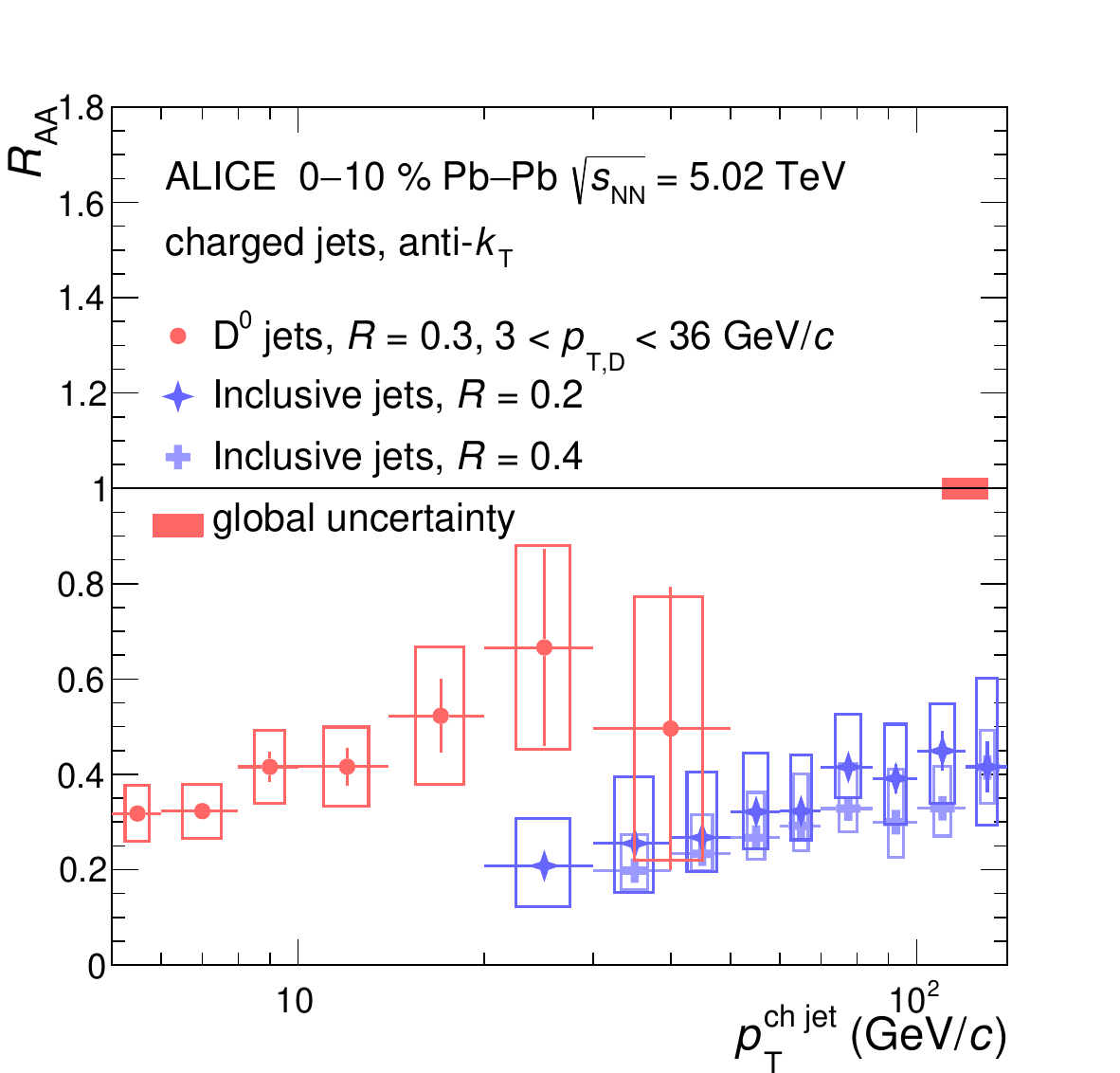}
\caption{Nuclear modification factor of \Dzero-tagged jets (red markers) and inclusive jets with resolution parameter $R=0.2$ (light blue) and $R=0.4$ (blue markers)~\cite{ALICE:2023waz}.
Statistical and systematic uncertainties are shown as vertical error bars and boxes, respectively. The global uncertainty originated from the $\left \langle T_\mathrm{AA} \right \rangle$ normalisation and the luminosity is displayed as a box around unity.
}
\label{fig:CharmJetRAA}
\end{figure}
The resulting \Raa of charm jets tagged with prompt \Dzero mesons is shown in Fig.~\ref{fig:CharmJetRAA}. The value of \Raa increases as a function of jet \pt. It varies from about 0.32 in the lowest jet-\pt interval to 0.5 in the highest jet-\pt\ interval, indicating a strong suppression of \Dzero-tagged-jet production in central (0--10\%) Pb--Pb collisions at $\snn\ = 5.02$~TeV.

Figure ~\ref{fig:CharmJetRAA} shows the comparison of  \Dzero-tagged-jet $R_{\rm AA}$ to the \Raa\ of inclusive charged-particle jets with $R=0.2$ and with $R=0.4$~\cite{ALICE:2023waz}. The two inclusive-jet \Raa measurements are compatible between each other within systematic uncertainties, with the \Raa for $R=0.4$ being systematically lower. The \Raa\ measurement for $R=0.3$ is not available, but it is reasonable to expect that it should lie between the two measurements.
Due to its broader kinematic range, only the measurement with $R=0.2$ will be considered from this point onward.
In the overlapping jet transverse momentum region ($20 < p_\mathrm{T}^\mathrm{ch~jet} < 50$~GeV/$c$), the inclusive jet $R_{\rm AA}$ is lower than the D$^{0}$-tagged-jet $R_{\rm AA}$, suggesting a reduced suppression of charm jets tagged with \Dzero mesons as compared to inclusive jet. The statistical significance of the difference is about 2.1$\sigma$ in the transverse momentum range $20~<~p_\mathrm{T}^\mathrm{ch~jet}~<~30$~GeV/$c$ and 1.8$\sigma$ in the entire transverse momentum region of the overlap between the two measurements.

In evaluating the comparison between inclusive jets and \Dzero-tagged jets, the steepness of the transverse-momentum differential yields in pp collisions, which were used as the reference for computing the \Raa, should be taken into account.
The momentum shift of inclusive and \Dzero-tagged jets due to the energy loss could differ due to the different steepness of their transverse-momentum spectrum and introduce a bias in the comparison of the two \Raa measurements.

\begin{figure}[tbp]
\centering
\includegraphics[width=0.6\textwidth]{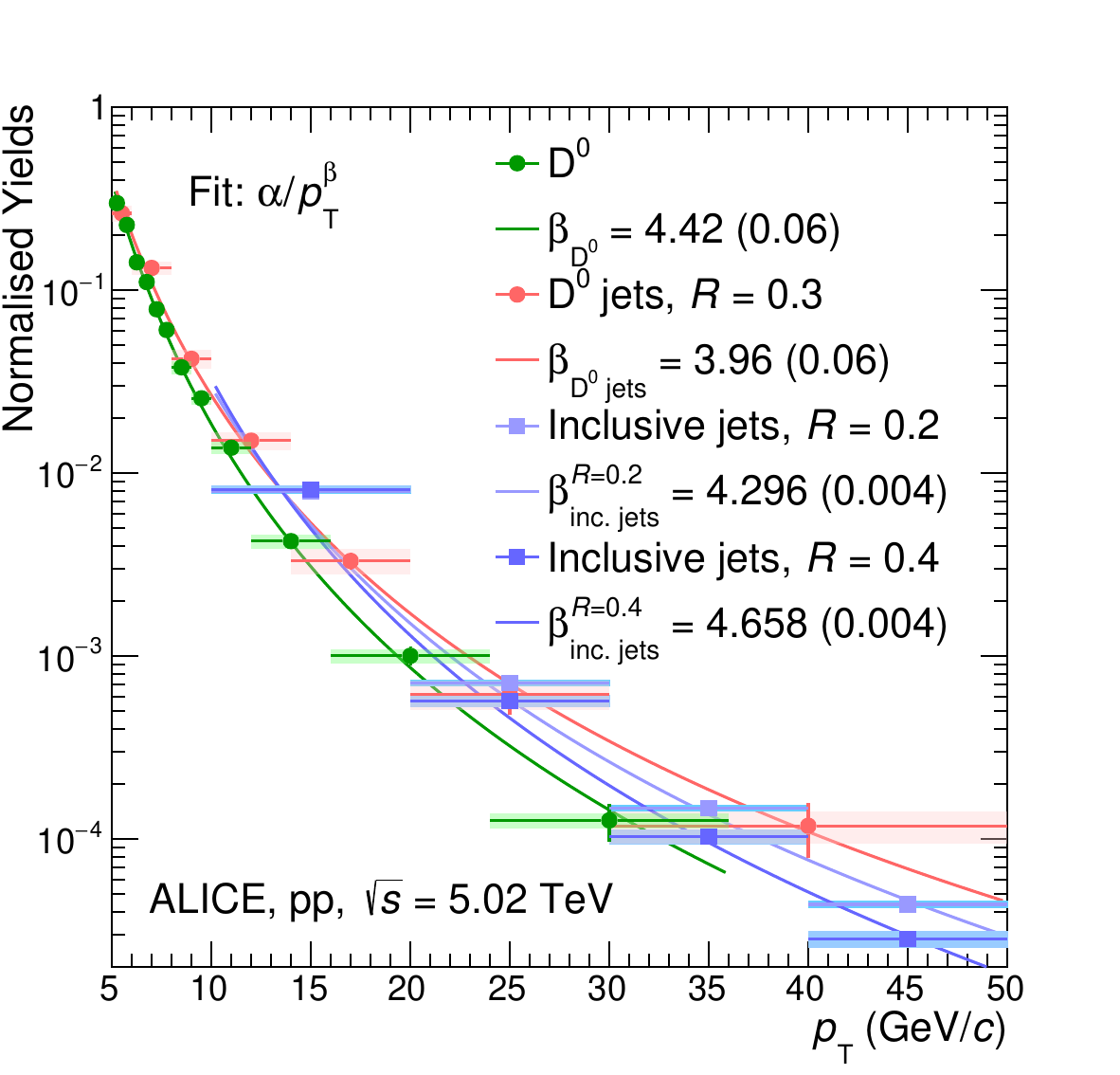}
\caption{Transverse momentum distributions of \Dzero mesons (green markers)~\cite{ALICE:2019nxm}, inclusive jets with jet resolution parameter 0.2 (light blue) and 0.4 (blue markers)~\cite{ALICE:2019wqv}, and \Dzero-tagged jets (red markers) in pp collisions at $\sqrt{s} = 5.02$ TeV (see text for details about the normalisation). Statistical and systematic uncertainties are shown as vertical error bars and boxes, respectively. The function $\alpha/p_{\rm{T}}^{\rm{\beta}}$ was fitted to each of the distributions (solid lines) to compare their steepness.}
\label{fig:PtSteepness}
\end{figure}

Figure~\ref{fig:PtSteepness} presents the \pt-differential yields of \Dzero mesons~\cite{ALICE:2019nxm}, \Dzero-tagged jets~\cite{ALICE:2022mur}, and inclusive jets with $R=0.2$ and $R=0.4$~\cite{ALICE:2019wqv} measured in pp collisions at $\sqrt{s} = 5.02$ TeV and normalised to their integral in given \pt intervals. The \Dzero-meson distribution was normalised in such a way that it has the same integral in the interval 5 to 6~\GeVc as the \Dzero-tagged jet distribution. The same was done for the inclusive-jet distributions in the interval 10 to 50~\GeVc.
Each transverse momentum distribution was fitted using a function $\alpha/p_{\rm{T}}^{\rm{\beta}}$, where $\alpha$ is a parameter that accounts for the normalisation of the distributions, and $\beta$ quantifies the steepness of the function. Both statistical and systematic uncertainties were considered for the fit.
For the \Dzero-meson and inclusive-jet transverse momentum spectra in pp collisions, it was found that $\mathrm{\beta}_{\rm{D}^{0}}=4.42\pm0.06$, $\beta_\mathrm{inc. jets}^{R = 0.2}=4.296\pm0.004$, and $\beta_\mathrm{inc. jets}^{R = 0.4}=4.468\pm0.004$, which show that these are steeper transverse momentum distributions in comparison to \Dzero-tagged jets with $\beta_\mathrm{\Dzero jets} = 3.96\pm0.06$. However, even assuming a hypothetical scenario where 80\% of the jets have a 10 \GeVc shift to lower \pt, the difference in the steepness between inclusive jets and jets tagged with a \Dzero could account for up to a 9\% difference between their nuclear modification factors, thus is not sufficiently large to explain the observed differences in their respective \Raa.
\begin{figure}[tbp]
\centering
\includegraphics[width=0.45\textwidth]{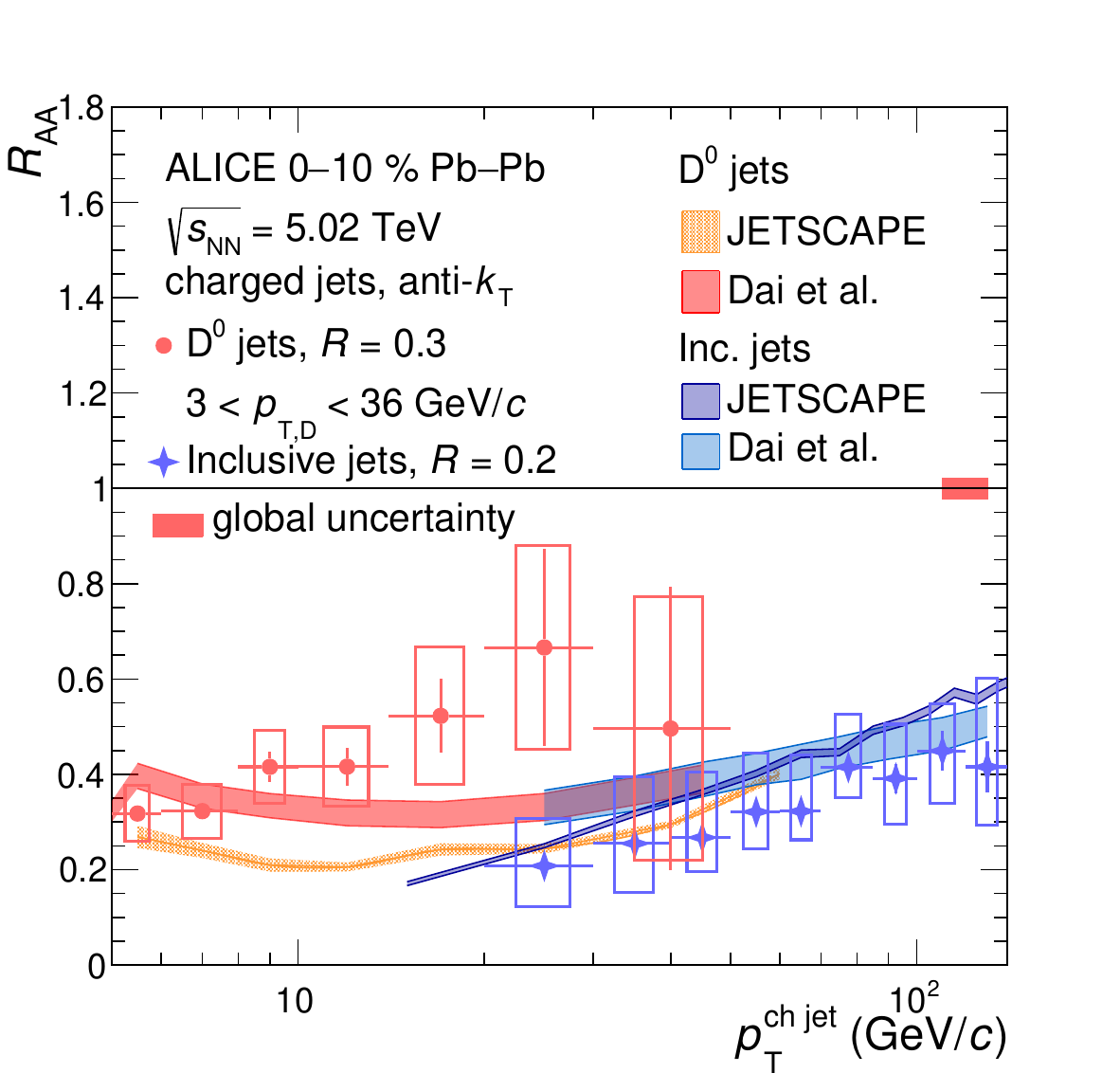} 
\includegraphics[width=0.45\textwidth]{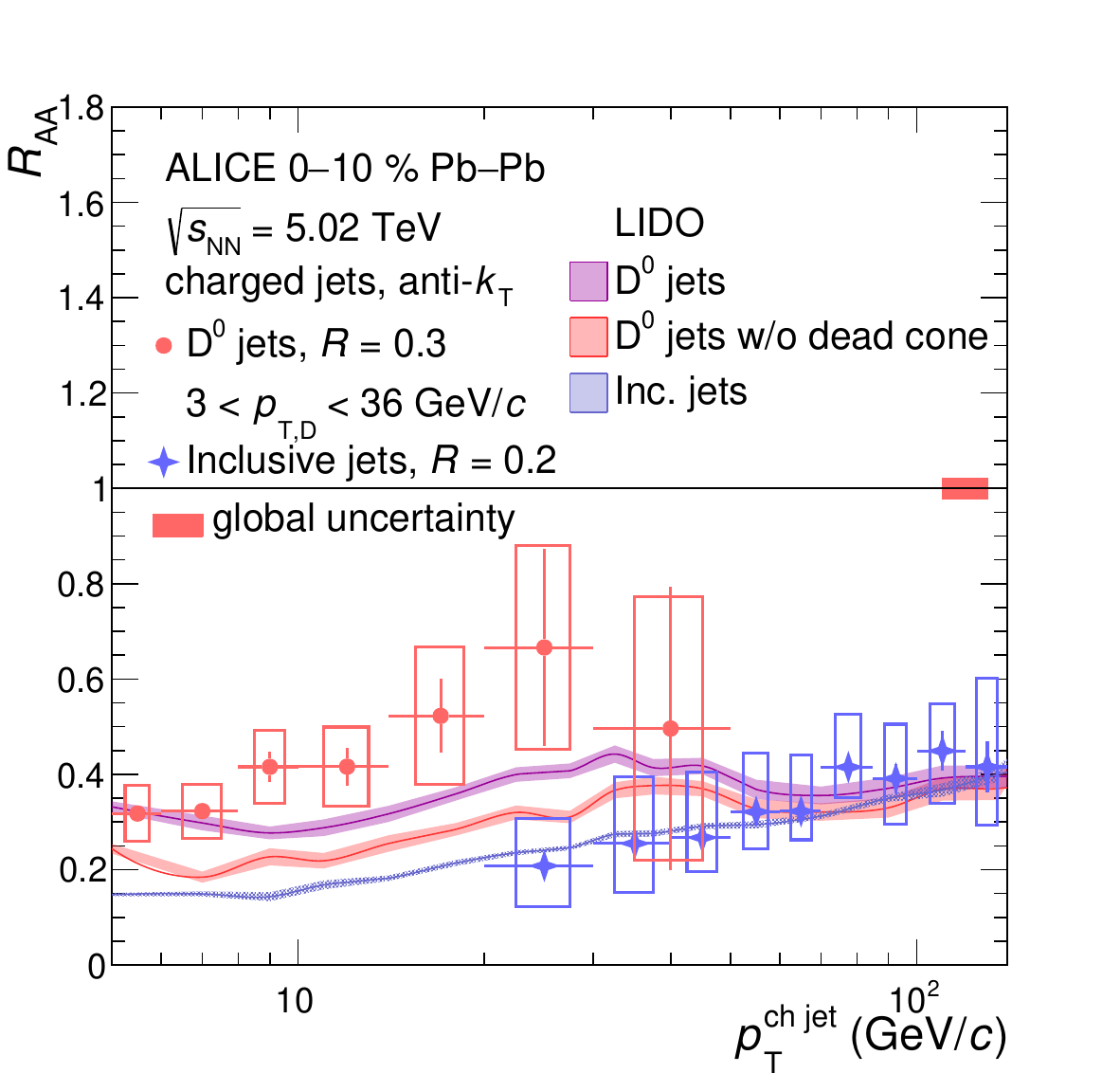}
\caption{
Left: comparison of the nuclear modification factor of \Dzero-tagged jets (and inclusive jets) with predictions from JETSCAPE~\cite{Putschke:2019yrg} and Dai et al.~\cite{Dai:2018mhw, Wang:2019xey, Wang:2021jgm}. The bands of the theory curves represent the systematic uncertainties on the model predictions.
Right: comparison to the LIDO~\cite{Ke:2018jem, Ke:2018tsh} predictions for \Dzero-tagged jets with and without dead-cone effect and predictions for inclusive jets.}
\label{fig:CharmJetRAAmodel}
\end{figure}

In Fig.~\ref{fig:CharmJetRAAmodel}, the nuclear modification factors of both \Dzero-tagged jets and inclusive jets are compared to calculations from different theory models, namely Dai et al.~\cite{Dai:2018mhw,Wang:2019xey, Wang:2021jgm}, JETSCAPE~\cite{Putschke:2019yrg} (left panel) and to LIDO~\cite{Ke:2018jem, Ke:2018tsh} model (right panel).

The calculation by Dai et al.~\cite{Dai:2018mhw, Wang:2019xey, Wang:2021jgm} is based on a Langevin transport model, which describes the evolution of heavy quarks and their collisional energy loss, as well as a higher-twist description of radiative energy loss for both heavy and light partons. This model also includes the dead-cone effect for heavy quarks and utilises a (2+1)-dimensional viscous hydrodynamic medium with averaged initial conditions. The model uses a fixed jet transport parameter (proportional to the local parton density in the medium) $\hat{q}_0 = 1.2$ GeV$^2$/fm and varies its value ($1<~\hat{q}_0~<1.5~$ GeV$^2$/fm) to estimate the uncertainties of the model predictions. 

The model provides an overall reasonable description of the nuclear modification factor (\Raa) for both inclusive and \Dzero-tagged jets. For inclusive jets, the predictions closely follow the data trend, although they tend to lie near the upper edge of the experimental systematic uncertainties. For \Dzero-tagged jets, the model lies slightly below the data for $p_\mathrm{T}^\mathrm{ch~jet} > 12$ \GeVc, but remains overall consistent with the measurements within approximately 2.6$\sigma$. In the $p_\mathrm{T}$ region where the two jet samples overlap, the model predicts a similar suppression pattern for inclusive and \Dzero-tagged jets.

The similarity between the suppression of inclusive and \Dzero-tagged jets seen in Dai et al. model predictions in the overlap region has been further investigated in a recent study~\cite{Wang:2024yag}, which employs a transport model based on next-to-leading order (NLO) calculations combined with vacuum parton showering. Two main factors are identified. First, \Dzero-tagged jets in the generated sample include a significant contribution from gluon splitting ($g \rightarrow \rm{ c\bar{c}}$), exceeding 40\% at $p_\mathrm{T}^\mathrm{ch~jet} > 30$~\GeVc. These gluon-initiated jets have a broader structure and lose more energy in the QGP than charm-quark-initiated jets, reducing the overall \Raa and bringing it closer to that of inclusive jets.
Second, the similarity in \Raa does not exclude the presence of mass-dependent energy loss effects. When charm-quark- and gluon-initiated jets are separated in the model, a clear hierarchy consistent with the expected mass ordering emerges. 

The JETSCAPE model~\cite{Putschke:2019yrg} includes a modification of the parton shower due to the interaction with the medium using MATTER~\cite{Cao:2017qpx} at high parton virtuality, and LBT~\cite{He:2015pra} at low parton virtuality. JETSCAPE underestimates the \Raa\ of \Dzero-tagged jets, while showing good agreement with the inclusive jet \Raa\ measurement.

LIDO~\cite{Ke:2018jem, Ke:2018tsh} is a partonic transport model that incorporates heavy-quark scatterings with medium partons using matrix elements calculated in perturbative QCD. It describes the transport of heavy quarks between scatterings through a Boltzmann type equation, including both elastic collisions and medium-induced parton radiation. The transport coefficients in LIDO are calibrated using a Bayesian analysis to match the measured single-inclusive nuclear modification factors of light-flavour hadrons and D mesons at the LHC and RHIC~\cite{Ke:2020clc}. The model introduces the parameter $\mu_\mathrm{min}$ to control the coupling between the jet and the medium, based on measurements of the $R_\mathrm{AA}$ of B and D mesons. The value of this parameter is proportional to the temperature of the QGP, $T$.
LIDO predictions, shown in the right panel of Fig.~\ref{fig:CharmJetRAAmodel}, were obtained with $\mu_\mathrm{min} = 1.8 \pi T$ and demonstrate good agreement with the data in terms of the shape and trend for both inclusive jets and \Dzero-tagged jets. 

Compared to the Dai et al. model, the LIDO framework, also based on next-to-leading-order event generation, predicts a more pronounced separation between the \Raa of inclusive and \Dzero-tagged jets in the overlapping $p_\mathrm{T}$ region, indicating that differences in the modeling of transport dynamics and jet-medium interactions might impact the predicted flavour dependence of jet quenching.

The LIDO calculation for inclusive jets down to $\pt=5$ \GeVc (compatible with the minimum \pt of the \Dzero-tagged jet measurement) predicts a larger suppression of inclusive jets also at low \pt. In the $p_\mathrm{T}^\mathrm{ch~jet} > 12$~\GeVc region, the model is compatible with the data within 2.2$\sigma$. Moreover, the result obtained without including the dead-cone effect in the calculations is also shown to single out the quark-mass effect in this model. Obtaining inclusive-jet measurements at even lower $p_{\mathrm{T}}$ would provide a crucial benchmark, as a direct comparison between inclusive and charm-tagged jets in this region would significantly improve our understanding of the charm in-medium energy-loss mechanism.

The comparison with the LIDO calculation without the dead-cone effect shows a larger suppression of the nuclear modification factor for charm jets at
$p_{\mathrm{T}}^{\mathrm{ch\,jet}} < 50~\mathrm{GeV}/c$, indicating that dead-cone effects play a significant role among the charm mass effects in this kinematic region. 
More precise measurements of \Dzero-tagged jets are necessary to better constrain the relative contribution of dead-cone effects compared to other mechanisms, such as mass-dependent fragmentation effects and differences in colour charge between partons. For $p_\mathrm{T}^\mathrm{ch~jet} >$ 50 \GeVc, the LIDO model predicts a similar behaviour for charm jets and inclusive jets. The latter behaviour is likely caused by a superposition of two effects: on one hand the constraints on the \Dzero-meson kinematics play a dominant role in shaping the observed trend and on the other hand the dead-cone effect becomes less relevant for increasing jet \pt.

The residual difference between the model curves of \Dzero-tagged jets without dead-cone effect and inclusive jets can possibly be attributed to the different Casimir factors of quarks and gluons, as the inclusive jet sample is a mixture of gluon- and light-quark-initiated jets.

\section{Summary}
\label{summary}
In this paper, the measurement of the production yield and nuclear modification factor of charm jets tagged with fully reconstructed \Dzero mesons in central (0--10\%) Pb--Pb collisions at a centre-of-mass energy per nucleon pair $\snn~=~5.02$ TeV has been reported. The use of \Dzero\ mesons for charm-jet tagging extends the measurement of the charm-jet production to transverse momentum values down to $p_\mathrm{T}^\mathrm{ch~jet} = 5$~\GeVc.

The nuclear modification factor of charm jets tagged with \Dzero mesons, which quantifies the suppression of the jet yield due to interactions with the medium, was measured for the first time. It was computed by comparing the \Dzero-tagged-jet yield in Pb--Pb collisions with the production cross section of \Dzero-tagged jets in pp collisions at the same collision energy as a reference and it was found to be suppressed in the full measured \pt\ range. The comparison with the nuclear modification factor of inclusive jets at the same centre-of-mass energy and centrality class indicates a lower suppression of charm jets compared to light-quark and gluon jets when traversing the medium, with a significance of about 2$\sigma$ in the transverse momentum range $20~<~p_\mathrm{T}^\mathrm{ch~jet}~<~50$~GeV/$c$. This difference can be attributed to the interplay of effects arising from the mass of the initial quark and the colour charge. The heavier quark mass suppresses the energy loss of charm quarks compared to light quarks, while the larger colour charge enhances the energy loss for gluon-initiated jets, which are more abundant at lower \pt, compared to quark-initiated jets.

The results are compared with JETSCAPE, Dai et al., and LIDO theoretical calculations of in-medium energy loss including quark-mass and colour-charge effects.
Among the models considered, LIDO qualitatively describes the ordering between charm jets and inclusive jets and it also shows the best quantitative agreement with the data. Specific modifications of the LIDO predictions indicate that dead-cone effect is an important factor in the suppression of radiative processes and it is a fundamental factor differentiating the nuclear modification factor of inclusive jets
from that of \Dzero-tagged jets in the kinematic region of the measurement.

The LIDO model also predicts a difference between the \Raa of inclusive jets and that of \Dzero-tagged jets without the dead-cone effect and it can be possibly attributed to the influence of colour charge effects.

Overall, this measurement provides important insights into the modification of charm jets in the QGP created in heavy-ion collisions and sheds light on the quark-mass and colour-charge dependence of the in-medium energy loss mechanisms.


\newenvironment{acknowledgement}{\relax}{\relax}
\begin{acknowledgement}
\section*{Acknowledgements}

The ALICE Collaboration would like to thank all its engineers and technicians for their invaluable contributions to the construction of the experiment and the CERN accelerator teams for the outstanding performance of the LHC complex.
The ALICE Collaboration gratefully acknowledges the resources and support provided by all Grid centres and the Worldwide LHC Computing Grid (WLCG) collaboration.
The ALICE Collaboration acknowledges the following funding agencies for their support in building and running the ALICE detector:
A. I. Alikhanyan National Science Laboratory (Yerevan Physics Institute) Foundation (ANSL), State Committee of Science and World Federation of Scientists (WFS), Armenia;
Austrian Academy of Sciences, Austrian Science Fund (FWF): [M 2467-N36] and Nationalstiftung f\"{u}r Forschung, Technologie und Entwicklung, Austria;
Ministry of Communications and High Technologies, National Nuclear Research Center, Azerbaijan;
Conselho Nacional de Desenvolvimento Cient\'{\i}fico e Tecnol\'{o}gico (CNPq), Financiadora de Estudos e Projetos (Finep), Funda\c{c}\~{a}o de Amparo \`{a} Pesquisa do Estado de S\~{a}o Paulo (FAPESP) and Universidade Federal do Rio Grande do Sul (UFRGS), Brazil;
Bulgarian Ministry of Education and Science, within the National Roadmap for Research Infrastructures 2020-2027 (object CERN), Bulgaria;
Ministry of Education of China (MOEC) , Ministry of Science \& Technology of China (MSTC) and National Natural Science Foundation of China (NSFC), China;
Ministry of Science and Education and Croatian Science Foundation, Croatia;
Centro de Aplicaciones Tecnol\'{o}gicas y Desarrollo Nuclear (CEADEN), Cubaenerg\'{\i}a, Cuba;
Ministry of Education, Youth and Sports of the Czech Republic, Czech Republic;
The Danish Council for Independent Research | Natural Sciences, the VILLUM FONDEN and Danish National Research Foundation (DNRF), Denmark;
Helsinki Institute of Physics (HIP), Finland;
Commissariat \`{a} l'Energie Atomique (CEA) and Institut National de Physique Nucl\'{e}aire et de Physique des Particules (IN2P3) and Centre National de la Recherche Scientifique (CNRS), France;
Bundesministerium f\"{u}r Forschung, Technologie und Raumfahrt (BMFTR) and GSI Helmholtzzentrum f\"{u}r Schwerionenforschung GmbH, Germany;
General Secretariat for Research and Technology, Ministry of Education, Research and Religions, Greece;
National Research, Development and Innovation Office, Hungary;
Department of Atomic Energy Government of India (DAE), Department of Science and Technology, Government of India (DST), University Grants Commission, Government of India (UGC) and Council of Scientific and Industrial Research (CSIR), India;
National Research and Innovation Agency - BRIN, Indonesia;
Istituto Nazionale di Fisica Nucleare (INFN), Italy;
Japanese Ministry of Education, Culture, Sports, Science and Technology (MEXT) and Japan Society for the Promotion of Science (JSPS) KAKENHI, Japan;
Consejo Nacional de Ciencia (CONACYT) y Tecnolog\'{i}a, through Fondo de Cooperaci\'{o}n Internacional en Ciencia y Tecnolog\'{i}a (FONCICYT) and Direcci\'{o}n General de Asuntos del Personal Academico (DGAPA), Mexico;
Nederlandse Organisatie voor Wetenschappelijk Onderzoek (NWO), Netherlands;
The Research Council of Norway, Norway;
Pontificia Universidad Cat\'{o}lica del Per\'{u}, Peru;
Ministry of Science and Higher Education, National Science Centre and WUT ID-UB, Poland;
Korea Institute of Science and Technology Information and National Research Foundation of Korea (NRF), Republic of Korea;
Ministry of Education and Scientific Research, Institute of Atomic Physics, Ministry of Research and Innovation and Institute of Atomic Physics and Universitatea Nationala de Stiinta si Tehnologie Politehnica Bucuresti, Romania;
Ministry of Education, Science, Research and Sport of the Slovak Republic, Slovakia;
National Research Foundation of South Africa, South Africa;
Swedish Research Council (VR) and Knut \& Alice Wallenberg Foundation (KAW), Sweden;
European Organization for Nuclear Research, Switzerland;
Suranaree University of Technology (SUT), National Science and Technology Development Agency (NSTDA) and National Science, Research and Innovation Fund (NSRF via PMU-B B05F650021), Thailand;
Turkish Energy, Nuclear and Mineral Research Agency (TENMAK), Turkey;
National Academy of  Sciences of Ukraine, Ukraine;
Science and Technology Facilities Council (STFC), United Kingdom;
National Science Foundation of the United States of America (NSF) and United States Department of Energy, Office of Nuclear Physics (DOE NP), United States of America.
In addition, individual groups or members have received support from:
Czech Science Foundation (grant no. 23-07499S), Czech Republic;
FORTE project, reg.\ no.\ CZ.02.01.01/00/22\_008/0004632, Czech Republic, co-funded by the European Union, Czech Republic;
European Research Council (grant no. 950692), European Union;
ICSC - Centro Nazionale di Ricerca in High Performance Computing, Big Data and Quantum Computing, European Union - NextGenerationEU;
Academy of Finland (Center of Excellence in Quark Matter) (grant nos. 346327, 346328), Finland.

\end{acknowledgement}

\bibliographystyle{utphys}   
\bibliography{bibliography}

\newpage
\appendix

%
%

\section{The ALICE Collaboration}
\label{app:collab}
\begin{flushleft} 
\small

S.~Acharya\,\orcidlink{0000-0002-9213-5329}\,$^{\rm 127}$, 
A.~Agarwal$^{\rm 135}$, 
G.~Aglieri Rinella\,\orcidlink{0000-0002-9611-3696}\,$^{\rm 32}$, 
L.~Aglietta\,\orcidlink{0009-0003-0763-6802}\,$^{\rm 24}$, 
M.~Agnello\,\orcidlink{0000-0002-0760-5075}\,$^{\rm 29}$, 
N.~Agrawal\,\orcidlink{0000-0003-0348-9836}\,$^{\rm 25}$, 
Z.~Ahammed\,\orcidlink{0000-0001-5241-7412}\,$^{\rm 135}$, 
S.~Ahmad\,\orcidlink{0000-0003-0497-5705}\,$^{\rm 15}$, 
S.U.~Ahn\,\orcidlink{0000-0001-8847-489X}\,$^{\rm 71}$, 
I.~Ahuja\,\orcidlink{0000-0002-4417-1392}\,$^{\rm 37}$
% ...

\section*{Affiliation Notes}

$^{\rm I}$ Deceased\\
$^{\rm II}$ Also at: Max-Planck-Institut fur Physik, Munich, Germany\\
$^{\rm III}$ Also at: Italian National Agency for New Technologies, Energy and Sustainable Economic Development (ENEA), Bologna, Italy\\
$^{\rm IV}$ Also at: Dipartimento DET del Politecnico di Torino, Turin, Italy\\
$^{\rm V}$ Also at: Yildiz Technical University, Istanbul, T\"{u}rkiye\\
$^{\rm VI}$ Also at: Department of Applied Physics, Aligarh Muslim University, Aligarh, India\\
$^{\rm VII}$ Also at: Institute of Theoretical Physics, University of Wroclaw, Poland\\
$^{\rm VIII}$ Also at: An institution covered by a cooperation agreement with CERN\\

\section*{Collaboration Institutes}

$^{1}$ A.I. Alikhanyan National Science Laboratory (Yerevan Physics Institute) Foundation, Yerevan, Armenia\\
$^{2}$ AGH University of Krakow, Cracow, Poland\\
$^{3}$ Bogolyubov Institute for Theoretical Physics, National Academy of Sciences of Ukraine, Kiev, Ukraine\\
$^{4}$ Bose Institute, Department of Physics  and Centre for Astroparticle Physics and Space Science (CAPSS), Kolkata, India\\
$^{5}$ California Polytechnic State University, San Luis Obispo, California, United States\\
$^{6}$ Central China Normal University, Wuhan, China\\
$^{7}$ Centro de Aplicaciones Tecnol\'{o}gicas y Desarrollo Nuclear (CEADEN), Havana, Cuba\\
$^{8}$ Centro de Investigaci\'{o}n y de Estudios Avanzados (CINVESTAV), Mexico City and M\'{e}rida, Mexico\\
$^{9}$ Chicago State University, Chicago, Illinois, United States\\
$^{10}$ China Institute of Atomic Energy, Beijing, China\\
$^{11}$ China University of Geosciences, Wuhan, China\\
$^{12}$ Chungbuk National University, Cheongju, Republic of Korea\\
$^{13}$ Comenius University Bratislava, Faculty of Mathematics, Physics and Informatics, Bratislava, Slovak Republic\\
$^{14}$ Creighton University, Omaha, Nebraska, United States\\
$^{15}$ Department of Physics, Aligarh Muslim University, Aligarh, India\\
$^{16}$ Department of Physics, Pusan National University, Pusan, Republic of Korea\\
$^{17}$ Department of Physics, Sejong University, Seoul, Republic of Korea\\
$^{18}$ Department of Physics, University of California, Berkeley, California, United States\\
$^{19}$ Department of Physics, University of Oslo, Oslo, Norway\\
$^{20}$ Department of Physics and Technology, University of Bergen, Bergen, Norway\\
$^{21}$ Dipartimento di Fisica, Universit\`{a} di Pavia, Pavia, Italy\\
$^{22}$ Dipartimento di Fisica dell'Universit\`{a} and Sezione INFN, Cagliari, Italy\\
$^{23}$ Dipartimento di Fisica dell'Universit\`{a} and Sezione INFN, Trieste, Italy\\
$^{24}$ Dipartimento di Fisica dell'Universit\`{a} and Sezione INFN, Turin, Italy\\
$^{25}$ Dipartimento di Fisica e Astronomia dell'Universit\`{a} and Sezione INFN, Bologna, Italy\\
$^{26}$ Dipartimento di Fisica e Astronomia dell'Universit\`{a} and Sezione INFN, Catania, Italy\\
$^{27}$ Dipartimento di Fisica e Astronomia dell'Universit\`{a} and Sezione INFN, Padova, Italy\\
$^{28}$ Dipartimento di Fisica `E.R.~Caianiello' dell'Universit\`{a} and Gruppo Collegato INFN, Salerno, Italy\\
$^{29}$ Dipartimento DISAT del Politecnico and Sezione INFN, Turin, Italy\\
$^{30}$ Dipartimento di Scienze MIFT, Universit\`{a} di Messina, Messina, Italy\\
$^{31}$ Dipartimento Interateneo di Fisica `M.~Merlin' and Sezione INFN, Bari, Italy\\
$^{32}$ European Organization for Nuclear Research (CERN), Geneva, Switzerland\\
$^{33}$ Faculty of Electrical Engineering, Mechanical Engineering and Naval Architecture, University of Split, Split, Croatia\\
$^{34}$ Faculty of Engineering and Science, Western Norway University of Applied Sciences, Bergen, Norway\\
$^{35}$ Faculty of Nuclear Sciences and Physical Engineering, Czech Technical University in Prague, Prague, Czech Republic\\
$^{36}$ Faculty of Physics, Sofia University, Sofia, Bulgaria\\
$^{37}$ Faculty of Science, P.J.~\v{S}af\'{a}rik University, Ko\v{s}ice, Slovak Republic\\
$^{38}$ Frankfurt Institute for Advanced Studies, Johann Wolfgang Goethe-Universit\"{a}t Frankfurt, Frankfurt, Germany\\
$^{39}$ Fudan University, Shanghai, China\\
$^{40}$ Gangneung-Wonju National University, Gangneung, Republic of Korea\\
$^{41}$ Gauhati University, Department of Physics, Guwahati, India\\
$^{42}$ Helmholtz-Institut f\"{u}r Strahlen- und Kernphysik, Rheinische Friedrich-Wilhelms-Universit\"{a}t Bonn, Bonn, Germany\\
$^{43}$ Helsinki Institute of Physics (HIP), Helsinki, Finland\\
$^{44}$ High Energy Physics Group,  Universidad Aut\'{o}noma de Puebla, Puebla, Mexico\\
$^{45}$ Horia Hulubei National Institute of Physics and Nuclear Engineering, Bucharest, Romania\\
$^{46}$ HUN-REN Wigner Research Centre for Physics, Budapest, Hungary\\
$^{47}$ Indian Institute of Technology Bombay (IIT), Mumbai, India\\
$^{48}$ Indian Institute of Technology Indore, Indore, India\\
$^{49}$ INFN, Laboratori Nazionali di Frascati, Frascati, Italy\\
$^{50}$ INFN, Sezione di Bari, Bari, Italy\\
$^{51}$ INFN, Sezione di Bologna, Bologna, Italy\\
$^{52}$ INFN, Sezione di Cagliari, Cagliari, Italy\\
$^{53}$ INFN, Sezione di Catania, Catania, Italy\\
$^{54}$ INFN, Sezione di Padova, Padova, Italy\\
$^{55}$ INFN, Sezione di Pavia, Pavia, Italy\\
$^{56}$ INFN, Sezione di Torino, Turin, Italy\\
$^{57}$ INFN, Sezione di Trieste, Trieste, Italy\\
$^{58}$ Inha University, Incheon, Republic of Korea\\
$^{59}$ Institute for Gravitational and Subatomic Physics (GRASP), Utrecht University/Nikhef, Utrecht, Netherlands\\
$^{60}$ Institute of Experimental Physics, Slovak Academy of Sciences, Ko\v{s}ice, Slovak Republic\\
$^{61}$ Institute of Physics, Homi Bhabha National Institute, Bhubaneswar, India\\
$^{62}$ Institute of Physics of the Czech Academy of Sciences, Prague, Czech Republic\\
$^{63}$ Institute of Space Science (ISS), Bucharest, Romania\\
$^{64}$ Institut f\"{u}r Kernphysik, Johann Wolfgang Goethe-Universit\"{a}t Frankfurt, Frankfurt, Germany\\
$^{65}$ Instituto de Ciencias Nucleares, Universidad Nacional Aut\'{o}noma de M\'{e}xico, Mexico City, Mexico\\
$^{66}$ Instituto de F\'{i}sica, Universidade Federal do Rio Grande do Sul (UFRGS), Porto Alegre, Brazil\\
$^{67}$ Instituto de F\'{\i}sica, Universidad Nacional Aut\'{o}noma de M\'{e}xico, Mexico City, Mexico\\
$^{68}$ iThemba LABS, National Research Foundation, Somerset West, South Africa\\
$^{69}$ Jeonbuk National University, Jeonju, Republic of Korea\\
$^{70}$ Johann-Wolfgang-Goethe Universit\"{a}t Frankfurt Institut f\"{u}r Informatik, Fachbereich Informatik und Mathematik, Frankfurt, Germany\\
$^{71}$ Korea Institute of Science and Technology Information, Daejeon, Republic of Korea\\
$^{72}$ KTO Karatay University, Konya, Turkey\\
$^{73}$ Laboratoire de Physique Subatomique et de Cosmologie, Universit\'{e} Grenoble-Alpes, CNRS-IN2P3, Grenoble, France\\
$^{74}$ Lawrence Berkeley National Laboratory, Berkeley, California, United States\\
$^{75}$ Lund University Department of Physics, Division of Particle Physics, Lund, Sweden\\
$^{76}$ Nagasaki Institute of Applied Science, Nagasaki, Japan\\
$^{77}$ Nara Women{'}s University (NWU), Nara, Japan\\
$^{78}$ National and Kapodistrian University of Athens, School of Science, Department of Physics , Athens, Greece\\
$^{79}$ National Centre for Nuclear Research, Warsaw, Poland\\
$^{80}$ National Institute of Science Education and Research, Homi Bhabha National Institute, Jatni, India\\
$^{81}$ National Nuclear Research Center, Baku, Azerbaijan\\
$^{82}$ National Research and Innovation Agency - BRIN, Jakarta, Indonesia\\
$^{83}$ Niels Bohr Institute, University of Copenhagen, Copenhagen, Denmark\\
$^{84}$ Nikhef, National institute for subatomic physics, Amsterdam, Netherlands\\
$^{85}$ Nuclear Physics Group, STFC Daresbury Laboratory, Daresbury, United Kingdom\\
$^{86}$ Nuclear Physics Institute of the Czech Academy of Sciences, Husinec-\v{R}e\v{z}, Czech Republic\\
$^{87}$ Oak Ridge National Laboratory, Oak Ridge, Tennessee, United States\\
$^{88}$ Ohio State University, Columbus, Ohio, United States\\
$^{89}$ Physics department, Faculty of science, University of Zagreb, Zagreb, Croatia\\
$^{90}$ Physics Department, Panjab University, Chandigarh, India\\
$^{91}$ Physics Department, University of Jammu, Jammu, India\\
$^{92}$ Physics Program and International Institute for Sustainability with Knotted Chiral Meta Matter (SKCM2), Hiroshima University, Hiroshima, Japan\\
$^{93}$ Physikalisches Institut, Eberhard-Karls-Universit\"{a}t T\"{u}bingen, T\"{u}bingen, Germany\\
$^{94}$ Physikalisches Institut, Ruprecht-Karls-Universit\"{a}t Heidelberg, Heidelberg, Germany\\
$^{95}$ Physik Department, Technische Universit\"{a}t M\"{u}nchen, Munich, Germany\\
$^{96}$ Politecnico di Bari and Sezione INFN, Bari, Italy\\
$^{97}$ Research Division and ExtreMe Matter Institute EMMI, GSI Helmholtzzentrum f\"ur Schwerionenforschung GmbH, Darmstadt, Germany\\
$^{98}$ Saga University, Saga, Japan\\
$^{99}$ Saha Institute of Nuclear Physics, Homi Bhabha National Institute, Kolkata, India\\
$^{100}$ School of Physics and Astronomy, University of Birmingham, Birmingham, United Kingdom\\
$^{101}$ Secci\'{o}n F\'{\i}sica, Departamento de Ciencias, Pontificia Universidad Cat\'{o}lica del Per\'{u}, Lima, Peru\\
$^{102}$ Stefan Meyer Institut f\"{u}r Subatomare Physik (SMI), Vienna, Austria\\
$^{103}$ SUBATECH, IMT Atlantique, Nantes Universit\'{e}, CNRS-IN2P3, Nantes, France\\
$^{104}$ Sungkyunkwan University, Suwon City, Republic of Korea\\
$^{105}$ Suranaree University of Technology, Nakhon Ratchasima, Thailand\\
$^{106}$ Technical University of Ko\v{s}ice, Ko\v{s}ice, Slovak Republic\\
$^{107}$ The Henryk Niewodniczanski Institute of Nuclear Physics, Polish Academy of Sciences, Cracow, Poland\\
$^{108}$ The University of Texas at Austin, Austin, Texas, United States\\
$^{109}$ Universidad Aut\'{o}noma de Sinaloa, Culiac\'{a}n, Mexico\\
$^{110}$ Universidade de S\~{a}o Paulo (USP), S\~{a}o Paulo, Brazil\\
$^{111}$ Universidade Estadual de Campinas (UNICAMP), Campinas, Brazil\\
$^{112}$ Universidade Federal do ABC, Santo Andre, Brazil\\
$^{113}$ Universitatea Nationala de Stiinta si Tehnologie Politehnica Bucuresti, Bucharest, Romania\\
$^{114}$ University of Cape Town, Cape Town, South Africa\\
$^{115}$ University of Derby, Derby, United Kingdom\\
$^{116}$ University of Houston, Houston, Texas, United States\\
$^{117}$ University of Jyv\"{a}skyl\"{a}, Jyv\"{a}skyl\"{a}, Finland\\
$^{118}$ University of Kansas, Lawrence, Kansas, United States\\
$^{119}$ University of Liverpool, Liverpool, United Kingdom\\
$^{120}$ University of Science and Technology of China, Hefei, China\\
$^{121}$ University of South-Eastern Norway, Kongsberg, Norway\\
$^{122}$ University of Tennessee, Knoxville, Tennessee, United States\\
$^{123}$ University of the Witwatersrand, Johannesburg, South Africa\\
$^{124}$ University of Tokyo, Tokyo, Japan\\
$^{125}$ University of Tsukuba, Tsukuba, Japan\\
$^{126}$ Universit\"{a}t M\"{u}nster, Institut f\"{u}r Kernphysik, M\"{u}nster, Germany\\
$^{127}$ Universit\'{e} Clermont Auvergne, CNRS/IN2P3, LPC, Clermont-Ferrand, France\\
$^{128}$ Universit\'{e} de Lyon, CNRS/IN2P3, Institut de Physique des 2 Infinis de Lyon, Lyon, France\\
$^{129}$ Universit\'{e} de Strasbourg, CNRS, IPHC UMR 7178, F-67000 Strasbourg, France, Strasbourg, France\\
$^{130}$ Universit\'{e} Paris-Saclay, Centre d'Etudes de Saclay (CEA), IRFU, D\'{e}partment de Physique Nucl\'{e}aire (DPhN), Saclay, France\\
$^{131}$ Universit\'{e}  Paris-Saclay, CNRS/IN2P3, IJCLab, Orsay, France\\
$^{132}$ Universit\`{a} degli Studi di Foggia, Foggia, Italy\\
$^{133}$ Universit\`{a} del Piemonte Orientale, Vercelli, Italy\\
$^{134}$ Universit\`{a} di Brescia, Brescia, Italy\\
$^{135}$ Variable Energy Cyclotron Centre, Homi Bhabha National Institute, Kolkata, India\\
$^{136}$ Warsaw University of Technology, Warsaw, Poland\\
$^{137}$ Wayne State University, Detroit, Michigan, United States\\
$^{138}$ Yale University, New Haven, Connecticut, United States\\
$^{139}$ Yonsei University, Seoul, Republic of Korea\\
$^{140}$ Affiliated with an institute covered by a cooperation agreement with CERN\\
$^{141}$ Affiliated with an international laboratory covered by a cooperation agreement with CERN.\\

\end{flushleft} 

\end{document}